\documentclass[onecollarge,natbib]{svjour2}
\bibpunct{[}{]}{;}{n}{}{,} % to get "[numbered]" references from natbib
\smartqed  % flush right qed marks, e.g. at end of proof
\usepackage{graphicx}

\usepackage{amssymb}
\usepackage{amsmath}
\usepackage{graphicx}
\usepackage{dcolumn}
\usepackage{bm}
\usepackage{epsfig}
\usepackage{amssymb}
\usepackage{color}
\usepackage{natbib}

\newcommand{\bal}{\begin{align}}
\newcommand{\eal}{\end{align}}
\newcommand{\beq}{\begin{eqnarray}}
\newcommand{\eeq}{\end{eqnarray}}
\newcommand{\nneeq}{\nonumber \end{eqnarray}}

\newcommand{\nn}{\nonumber \\}
\newcommand{\es}{& = &}
\newcommand{\rs}{\, = \,}

\newcommand{\cH}{ {\cal H} }

\newcommand{\cG}{ {\cal G} }
\newcommand{\cT}{ {\cal T} }

\newcommand{\cU}{ {\cal U} }

\journalname{Few-Body Systems}
\begin{document}

\title{ Asymptocic Freedom of Gluons in Hamiltonian Dynamics
\thanks{Presented by Mar\'ia G\'omez-Rocha at Light Cone 2015, 21-25 September, 2015, INFN Frascati National Laboratories.}
}


\author{Mar\'ia G\'omez-Rocha     \and
        Stanis\l aw D. G\l azek %etc.
}


\institute{Mar\'ia G\'omez-Rocha \at
              ECT*,  Strada delle Tabarelle, 286,  I-38123 Villazzano Trento, Italy \\
              \email{mariagomezrocha@gmail.com}           %  \\
           \and
           Stanis\l aw D. G\l azek \at
              University of Warsaw, 
Pasteura 5, 02-093 Warsaw, Poland}

\date{Received: date / Accepted: date}

\maketitle

\begin{abstract}
We derive asymptotic freedom of gluons in terms of the renormalized $SU(3)$ Yang-Mills Hamiltonian in the Fock space.
Namely, we use the renormalization group procedure for effective particles (RGPEP) to calculate the three-gluon interaction term in the front-form Yang-Mills Hamiltonian using a perturbative expansion in powers of $g$ up to third order. The resulting three-gluon vertex is a function of the scale parameter $s$ that has an interpretation of the size of effective gluons.
The corresponding Hamiltonian running coupling constant
exhibits asymptotic freedom, and the corresponding Hamiltonian $\beta$-function coincides with the one  obtained in an earlier calculation using a different generator. 

\keywords{Asymptotic freedom \and Quantum chromodynamics \and Renormalization group \and  Relativistic Hamiltonian dynamics}
\end{abstract}

%%%%%%%%%%%%%%%%%%%%%%%%%%%%%%%%%%%
\section{Introduction}
\label{intro}
%%%%%%%%%%%%%%%%%%%%%%%%%%%%%%%%%%%

The renormalization group procedure for effective particles (RGPEP) has been developed during the last years~\cite{Glazek:2012qj,GlazekTrawinskiAdS,Gomez-Rocha:2015esa} as a non-perturbative tool for constructing bound-states in quantum chromodynamics (QCD)~\citep{Wilsonetalweakcoupling}.
It introduces the concept of effective particles, which differ from the bare or canonical ones, by having size $s$, corresponding to the momentum scale $\lambda=1/s$. 
Creation and annihilation of effective particles in the Fock space are described by the action of effective particle operators, $a^\dagger_s$ and $a_s$, on states built from the vacuum state $|0\rangle$ using $a_s^\dagger$; bare particle operators, $a^\dagger_0$ and $a_0$, appearing in the canonical Hamiltonian,  create and annihilate pointlike particles (with size $s=0$).

We are interested in calculating the evolution of quark and gluon quantum states, describing their dynamics and studying their binding. In a single formulation, the sought effective Hamiltonian must provide a means for the constituent-like behavior of quarks and gluons in hadrons with the measured quantum numbers, and also an explanation for the short-distance phenomena of  weakly interacting pointlike partons.

In this work, we apply the RGPEP to  interacting gluons in the absence of quarks. We will demonstrate that the RGPEP passes the test of describing asymptotic freedom, which is a precondition for any approach aiming at using QCD, especially for tackling nonperturbative issues, such as the ones that emerge when one allows effective gluons to have masses~\cite{Wilsonetalweakcoupling}.

We start from the regularized canonical Hamiltonian for quantum Yang-Mills field in the Fock space, obtained from the corresponding Lagrangian density. We use the RGPEP to  introduce effective particles and calculate a family of  effective Hamiltonians characterized by a scale or size parameter $s$. These Fock-space Hamiltonians depend on the effective-particle size parameter in an asymptotically free way: the coupling constant in a three-gluon interaction term vanishes with the inverse of $\ln(1/s)$.

In the following, we summarize the procedure, which is general and can be applied to any other quantum field theory.
For a more extended and detailed explanation, we refer the reader to Ref.~\cite{Gomez-Rocha:2015esa}.

%%%%%%%%%%%%%%%%%%%%%%%%%%%%%%%%%%%%%%%%%
\section{Renormalization group procedure for effective particles }
\label{sec:1}
%%%%%%%%%%%%%%%%%%%%%%%%%%%%%%%%%%%%%%%%%

%%%%%%%%%%%
\subsection{Initial Hamiltonian}
\label{subsec:2:1}
%%%%%%%%%%%

We derive the canonical Hamiltonian for Yang-Mills theories from the Lagrangian density:
\begin{eqnarray}
 {\cal L} = - {1 \over 2} \text{tr} \, F^{\mu \nu}
F_{\mu \nu} \ ,
\end{eqnarray}
where 
$F^{\mu \nu} = \partial^\mu A^\nu 
- \partial^\nu A^\mu + i g [A^\mu, A^\nu]$,  
$A^\mu = A^{a \mu} t^a$, 
$ [t^a,t^b] = i f^{abc} t^c$, which leads to the energy-momentum tensor, 
\beq
\cT^{\mu \nu} \es
-F^{a \mu \alpha} \partial^\nu A^a_\alpha + g^{\mu \nu} F^{a \alpha
\beta} F^a_{\alpha \beta}/4 \ .
\eeq
We choose the front-form (FF) of dynamics~\citep{Dirac1949} which consists of setting the quantization surface on the hyperplane $x^+=x^0+x^3=0$. Using  the gauge $A^+=0$, the Lagrange equations lead to the condition
\begin{eqnarray}
A^- \es 
{ 1 \over \partial^+ } \, 2 \, \partial^\perp A^\perp 
- { 2 \over \partial^{ + \, 2} } \ 
ig \, [ \partial^+ A^\perp, A^\perp] \ ,
\end{eqnarray}
so that the only degrees of freedom are the fields $A^\perp$.

Integration of $\cT^{+-}$ over the front $x^+=0$ leads to the FF energy of the constrained gluon field:
\begin{equation}
P^- = {1 \over 2}\int dx^- d^2 x^\perp \cT^{+-}\, |_{x^+=0} \quad .
\label{Hg}
\end{equation}
This operator contains a series of products of 2nd, 3rd or 4th powers of the field $A^\mu$ or their derivatives. The energy momentum tensor $\cT^{+-}$ can be written as
\beq
\cT^{+ -} = {\cal H}_{A^2} + {\cal H}_{A^3} + {\cal H}_{A^4} + {\cal
H}_{[\partial A A]^2} \ ,
\eeq
where
\cite{Casher:1976ae,Thorn:1979gv,BrodskyLepage}
\beq
\label{HA2}
{\cal H}_{A^2} \es - {1\over 2} A^{\perp a } (\partial^\perp)^2 A^{\perp a} \  , \\
\label{HA3}
{\cal H}_{A^3} \es  g \, i\partial_\alpha A_\beta^a [A^\alpha,A^\beta]^a  \ , \\
\label{HA4}
{\cal H}_{A^4} \es  - {1\over 4} g^2 \, [A_\alpha,A_\beta]^a[A^\alpha,A^\beta]^a  \ , \\
\label{HA2A2}
{\cal H}_{[\partial A A]^2} \es  {1\over 2}g^2 \,
[i\partial^+A^\perp,A^\perp]^a {1 \over (i\partial^+)^2 }
[i\partial^+A^\perp,A^\perp]^a \ .
\eeq
The quantum canonical Hamiltonian is obtained by replacing the field $A^\mu$ by
the quantum field operator
\beq
\hat A^\mu \es \sum_{\sigma c} \int [k] \left[ t^c \varepsilon^\mu_{k\sigma}
a_{k\sigma c} e^{-ikx} + t^c \varepsilon^{\mu *}_{k\sigma}
a^\dagger_{k\sigma c} e^{ikx}\right]_{x^+=0} \ ,
\eeq
where $[k] = \theta(k^+)
dk^+ d^2 k^\perp/(16\pi^3 k^+)$,
the polarization four-vector is defined as $\varepsilon^\mu_{k\sigma} 
= (\varepsilon^+_{k\sigma}=0, \varepsilon^-_{k\sigma} 
= 2k^\perp \varepsilon^\perp_\sigma/k^+, 
\varepsilon^\perp_\sigma)$ and the indices $\sigma$ and $c$ denote spin and color quantum numbers, respectively. The creation and annihilation operators  satisfy the commutation relations
\beq
\left[ a_{k\sigma c}, a^\dagger_{k'\sigma' c'} \right] 
\es 
k^+
\tilde \delta(k - k') \,\, \delta^{\sigma \sigma'}
\, \delta^{c c'} \ , 
\quad
\left[ a_{k\sigma c}, a_{k'\sigma' c'} \right] 
\ = \
\left[ a^\dagger_{k\sigma c}, a^\dagger_{k'\sigma' c'} \right] 
\ = \
0 \ ,
\eeq
with $\tilde \delta(p) = 16 \pi^3 \delta(p^+) \delta(p^1)
\delta(p^2)$.

The canonical Hamiltonian is divergent and  needs regularization. At every interaction term, every creation and annihilation operator in the canonical Hamiltonian is multiplied by a regulating factor\footnote{Other regulating functions are available~\cite{Gomez-Rocha:2015esa}. Finite dependence of the effective Hamiltonian on the small-x regularization may be thought to be related to the vacuum state problem, the phenomena of symmetry breaking and confinement~\cite{Wilsonetalweakcoupling}.}
\beq
r_{\Delta \delta}(\kappa^\perp, x) 
\es 
\exp(-\kappa^\perp / \Delta)\, x^\delta \theta(x-\epsilon) \ ,
\eeq 
where $x$ is the relative momentum fraction $x_{p/P}=p^+/P^+$, $\kappa$ is the relative transverse momentum, $\kappa_{p/P}=p^\perp-xP^\perp$, and $P$ is the total momentum in the term one considers.
The regulating function prevents the interaction terms from acting if the change of transverse momentum between gluons were to exceed $\Delta$, or if the change of longitudinal momentum fraction $x$ were to be smaller than  $\delta$.

\subsection{Derivation of the effective Hamiltonian}
The RGPEP transforms bare, or point-like creation and annihilation operators into effective ones~\citep{Glazek:2012qj}. Effective particle operators of size $s=t^{1/4}$ are related to bare ones by certain unitary transformation
\beq
\label{at}
a_t \es \cU_t \, a_0 \, \cU_t^\dagger  \ .
\eeq
The fact that the Hamiltonian operator cannot be affected by this change requires,
\beq
\cH_t(a_t) \es \cH_0(a_0)  \ , 
\eeq
which is equivalent to writing:
\beq
\label{cHt}
\cH_t(a_0) = \cU_t^\dagger \cH_0(a_0)\cU_t  \ .
\eeq
Differentiating both sides of~(\ref{cHt}) leads to the RGPEP equation:
\beq 
\label{ht1}
\cH'_t(a_0) \es
[ \cG_t(a_0) , \cH_t(a_0) ] \ ,
\eeq 
where $\cG_t = - \cU_t^\dagger \cU'_t$,
and therefore, 
$
\cU_t 
= 
T \exp{ \left( - \int_0^t d\tau \, \cG_\tau
\right) }$. $T$ denotes ordering in $\tau$.
The RGPEP equation~(\ref{ht1}) is the engine of this procedure. It governs the evolution of effective particles with the scale parameter $t$. It encodes the relation between pointlike quantum gluons appearing in the canonical Hamiltonian and the effective,
or constituent ones referred to by effective phenomenological models describing bound states.

 We choose the generator to be the commutator $\cG_t = [ \cH_f, \cH_{Pt} ]$,\footnote{Other generators are also allowed but may lead to more complicated expressions~\cite{Glazek:2000dc}. Our choice is similar to Wegner's~\cite{Wegner}.} where $\cH_f$ is the non-interacting term of the Hamiltonian and $\cH_{Pt}$ is defined in terms of $\cH_t$ . 

$\cH_t$ is a series of normal-ordered products of creation and annihilation operators,
\beq
\label{Hstructure} 
\cH_t(a_0) =
\sum_{n=2}^\infty \, 
\sum_{i_1, i_2, ..., i_n} \, c_t(i_1,...,i_n) \, \, a^\dagger_{0i_1}
\cdot \cdot \cdot a_{0i_n} \, .
\eeq 
$\cH_{Pt}$ differs from $\cH_t$ by 
 the vertex total $+$-momentum factor,
\beq
\label{HPstructure} 
\cH_{Pt}(a_0) \es
\sum_{n=2}^\infty \, 
\sum_{i_1, i_2, ..., i_n} \, c_t(i_1,...,i_n) \, 
\left( {1 \over
2}\sum_{k=1}^n p_{i_k}^+ \right)^2 \, \, a^\dagger_{0i_1}
\cdot \cdot \cdot a_{0i_n} \, .
\eeq 
The initial condition for the differential equation~(\ref{ht1}) is given by the regularized canonical Hamiltonian given in Section~\ref{subsec:2:1} plus counterterms.
More precisely, the initial condition is given by the physical fact that at very small distances or very high energies, the regularized canonical Hamiltonian must be recovered and any regularization dependence must be removed. 

We solve the RGPEP equation~(\ref{ht1}) for the effective Hamiltonians using an expansion in powers of the coupling constant $g$  up to third order and we focus our studies on the structure of the three-gluon term~\citep{Glazek:2012qj,Gomez-Rocha:2015esa}.

%%%%%%%%%%%%%%%%%%%%%%%%%%%%%%%%%%%
\section{The three-gluon vertex}
\label{sec:2}
%%%%%%%%%%%%%%%%%%%%%%%%%%%%%%%%%%%

The third-order effective Hamiltonian  expansion have the following structure:
\beq
\label{Hpert}
H_t \es 
H_{11,0,t} + H_{11,g^2,t} + H_{21,g,t} + 
H_{12,g,t} + H_{31,g^2,t} + H_{13,g^2,t} + 
H_{22,g^2,t} + H_{21,g^3,t} + H_{12,g^3,t}  \ .
\eeq
The first and second subscripts indicate the number of creation and annihilation operators, respectively. The third subscript labels the order in powers of $g$. Finally, the last label indicates the dependence on the scale parameter $t$.

The initial condition at $t=0$ has the form:
\beq
\label{Hper0}
H_0 \es 
H_{11,0,0} + H_{11,g^2,0} + H_{21,g,0} + 
H_{12,g,0} + H_{31,g^2,0} + H_{13,g^2,0} + 
H_{22,g^2,0} + H_{21,g^3,0} + H_{12,g^3,0}  \ ,
\eeq 
and consists of the regularized canonical Hamiltonian plus counterterms. The latter are calculated in such a way that $H_t$ remains finite when $\Delta\to\infty$. It is not possible to remove the small-$x$ cutoff $\delta$ at these point. However, this dependence will be of interest in higher-order calculations, since the small-$x$ phenomena are thought to be related to the vacuum-state behavior. The last step in the RGPEP is to replace bare creation and annihilation operators by effective ones.

\begin{figure}[h]
 \includegraphics[width=0.9\textwidth]{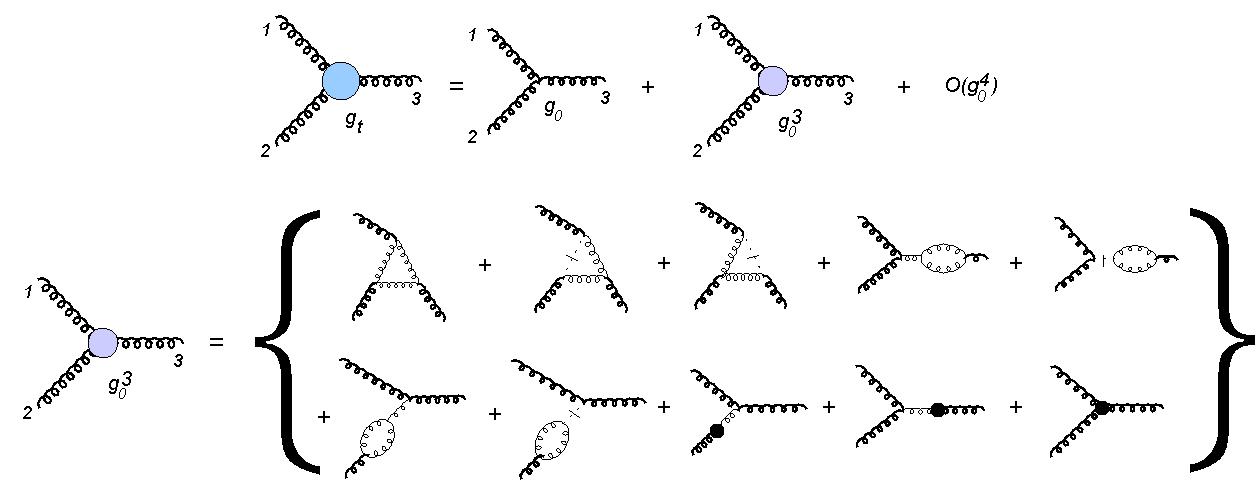}
 \caption{Graphical representation of terms contributing to the effective three-gluon vertex (third-order expansion)~\citep{Gomez-Rocha:2015esa}. Thin internal lines correspond to intermediate bare gluons and thick external lines  correspond to the creation and
annihilation operators that appear in the three-gluon FF Hamiltonian interaction
term for effective gluons of size $s$.
Dashed lines with transverse bars represent the combined contributions of terms~(\ref{HA4}) and~(\ref{HA2A2}). The black dots indicate counterterms.}
 \label{Fig-runingg}
\end{figure}

The three-gluon vertex and the running of the Hamiltonian coupling are encoded in the sum of first- and third-order terms:
\begin{eqnarray}
H_{(1+3),t} \es (H_{21,g,t} + 
H_{12,g,t}) + ( H_{21,g^3,t} + H_{12,g^3,t} ) \ .
\end{eqnarray}
The third-order solution requires the knowledge of the first and second-order solutions. The sum of all these contributions has the form\footnote{The subscripts 1,2,3 refer to the gluon lines indicated in Fig.~\ref{Fig-runingg}. So, e.g. $\kappa_{12}^\perp= x_{2/3}\kappa_{1/3}^\perp - x_{1/3}\kappa_{2/3}^\perp$ . }~\citep{Gomez-Rocha:2015esa} (see Fig.~\ref{Fig-runingg}):
\begin{eqnarray}
H_{(1+3),t} \es
\sum_{123}\int[123] \  \tilde \delta(k_1+k_2-k_3) 
\ f_{12t} \, \left[ \tilde Y_{21\,t}(x_1,\kappa_{12}^\perp, \sigma)a^\dagger_{1t} a^\dagger_{2t} a_{3t}  \ + \tilde Y_{12\,t}(x_1,\kappa_{12}^\perp, \sigma)\ a_{3t}^\dagger a_{2t} a_{1t} \right]\nn 
\end{eqnarray}
where $f_{12 t}=e^{-(k_1+k_2)^4t}$ is a form factor and $\tilde Y_{21\,t}(x_1,\kappa_{12}^\perp, \sigma)$ is the object of our study. 
We define the Hamiltonian coupling constant $g_t$ as the coefficient in front of the canonical color, spin and momentum dependent factor $Y_{123}(x_1,\kappa_{12}^\perp,\sigma)= i f^{c_1 c_2 c_3} [ \varepsilon_1^*\varepsilon_2^*
\cdot \varepsilon_3\kappa_{12}^\perp - \varepsilon_1^*\varepsilon_3 \cdot
\varepsilon_2^*\kappa_{12}^\perp {1\over x_{2/3}} - \varepsilon_2^*\varepsilon_3
\cdot \varepsilon_1^*\kappa_{12}^\perp {1\over x_{1/3}} ]$ in the limit $\kappa_{12}^\perp\to 0$, for some value of $x_1$ denoted by $x_0$.
So,
\beq
\label{consider}
\lim_{\kappa_{12}^\perp \to 0}
\tilde Y_t(x_1,\kappa_{12}^\perp, \sigma) 
\es
\lim_{\kappa_{12}^\perp \to 0}
\left[
c_t(x_1,\kappa_{12}^\perp) 
Y_{123}(x_1,\kappa_{12}^\perp, \sigma) 
+ 
g^3 \tilde T_{3 \,\text{finite}}(x_1,\kappa_{12}^\perp, \sigma) 
\right]  .
\eeq
where $\tilde T_{3 \,\text{finite}}(x_1,\kappa_{12}^\perp, \sigma)$ is a finite part contained in the counterterm and does not contribute to the running coupling,
\beq
\lim_{\kappa_{12}^\perp \to 0}
c_t(x_1,\kappa_{12}^\perp) 
\es
g + g^3 \lim_{\kappa_{12}^\perp \to 0}
\left[
  c_{3t  }(x_1,\kappa_{12}^\perp) 
- 
  c_{3t_0}(x_1,\kappa_{12}^\perp) 
\right] \ . 
  \label{limitc3}
\eeq
And assuming some value for $g_0$ at some small $t_0$, $g_{t_0}=g_0$,
\beq
\label{gl1}
g_t
& \equiv &  c_t(x_1) 
\ = \
\label{gl2}
g_0 + g_0^3 
\left[
  c_{3t  }(x_1) 
- 
  c_{3t_0}(x_1) 
\right] \ .
\eeq
We introduce now the momentum scale parameter $\lambda=t^{-1/4}$. This yields,
\beq
\label{gl}
g_\lambda \es
g_0 - { g_0^3 \over 48 \pi^2 }   N_c \,   11 \,\ln
{ \lambda \over \lambda_0} \ .
\eeq 
Differentiation of the latter with respect to $\lambda$ leads to
\beq
\lambda {d \over d\lambda} \, g_\lambda
\es \beta_0 g_\lambda^3 \ , \quad \text{with}\quad \beta_0 \rs - { 11 N_c \over 48\pi^2 } \ .
\eeq
This result equals the asymptotic freedom result 
in Refs.~\cite{Gross:1973id,Politzer:1973fx},
when one identifies $\lambda$ with the momentum 
scale of external gluon lines in Feynman diagrams.
Our result also coincides with the expression obtained in~\citep{Glazek:2000dc}, where an analogous calculation were performed using a different generator.

\section{Summary and conclusion}

We have applied the RGPEP to the quantum $SU(3)$ Yang-Mills theory and extracted the running coupling from the three-gluon-vertex  term in the third-order effective Hamiltonian.
The result turns out to be independent of the choice of the generator, as it coincides with the one obtained in an analogous calculation performed in~\cite{Glazek:2000dc}, using a different generator. The present generator, however, leads to simpler equations than the older one, which is desired and needed for our forthcoming forth-order calculations, required for any attempt at description of physical systems using QCD~\citep{Wilsonetalweakcoupling}. 
The obtained running coupling is of the form that is familiar from other formalism and renormalization schemes and passes the test of producing asymptotic freedom, which any method aiming at solving QCD must past.

\begin{acknowledgements}
Part of this work was supported by the Austrian Science Fund (FWF) under project No. P25121-N27. Fig.~\ref{Fig-runingg} was produced with JaxoDraw~\citep{Binosi:2003yf}. 
\end{acknowledgements}

% BibTeX users please use
%\bibliographystyle{spbasic}

\bibliography{bibRGPEP}

\bibliographystyle{ieeetr}

\end{document}